\documentclass[11pt,titlepage]{article}
\usepackage{psfrag,epsfig,amsfonts,amssymb,amsmath,graphicx}

\usepackage[round,numbers,sort&compress]{natbib} 
\usepackage[nofiglist,nomarkers]{endfloat}

\newcommand{\NL}{N_{\mathrm{lin}}}

\newcommand{\NC}{N_{\mathrm{clu}}}
\newcommand{\RC}{R_{\mathrm{clu}}}
\newcommand{\rL}{\mathbf{r}_{i}^{\mathrm{(lin)}}}
\newcommand{\rR}{\mathbf{r}_{i}^{\mathrm{(rec)}}}

\begin{document}

\title{Hidden multiple bond effects in dynamic force spectroscopy}

\author{Sebastian Getfert\thanks{
           	Corresponding author. Address: Fakult\"at f\"ur Physik,
           	Universit\"at Bielefeld, Universit\"atsstr. 25, 33615,
           	Bielefeld, Germany.  E-Mail: Getfert@Physik.Uni-Bielefeld.De}
           	\hspace*{0.2cm} and \hspace*{0.2cm}
        	Peter Reimann\\
       	 Universit\"at Bielefeld, Fakult\"at f\"ur Physik, 33615 Bielefeld, Germany}

% Revision date - uncomment to exclude date in the final version
\date{}

% Running head
\pagestyle{myheadings}

% generate the title page from the info in the headers above
\maketitle

\begin{abstract}
In dynamic force spectroscopy, a (bio-)molecular
complex is subjected
to a steadily increasing force until the 
chemical bond breaks.
Repeating the same experiment many times results in a 
broad distribution of rupture forces,
whose quantitative interpretation
represents a formidable theoretical challenge.
In this study we address the situation
that more than a single molecular bond is involved
in one experimental run, giving rise to multiple rupture 
events which are even more difficult to analyze 
and thus are usually eliminated as
far as possible from the further evaluation 
of the experimental data.
We develop and numerically solve a detailed model of a complete 
dynamic force spectroscopy experiment including 
a possible clustering of molecules on the substrate surface, 
the formation of bonds, 
their dissociation under load, and the post 
processing of the force extension curves.
We show that the data, remaining after elimination of 
obvious multiple rupture events,  may still contain a 
considerable number of ``hidden'' multiple bonds, 
which are experimentally indistinguishable from ``true'' 
single bonds, but which have considerable effects on 
the resulting rupture force statistics and its 
consistent theoretical interpretation.
\end{abstract}

\maketitle
%%%%%%%%%%%%%%%%%%%%%%%%%%%%%%%%%%%%%%%%%%%%%%%%%%%%%%%%%%%%%%
\section{Introduction}
\label{introduction}
The specific binding of a ligand molecule to a receptor 
protein is an essential functional principle of molecular 
recognition and information processing 
in biological systems, e.g. in the context of
genome replication and transcription, 
enzymatic activity, initiation of infection
and  immune response \cite{hin06}.
Furthermore, their force-induced dissociation is 
of great interest with respect to adhesion and 
cohesion of any type of biological matter
and their selectivity can be exploited in various 
bioanalytical and biomedical devices \cite{mer01,rit06}.
The invention and continuous refinement 
of atomic force microscopy (AFM) 
and other related techniques
\cite{mer01,rit06} has led to a tremendous 
progress in our abilities to directly investigate such 
processes on the molecular level 
\cite{flo94,dam96}.
Accordingly, dynamic force spectroscopy 
has become a standard tool in experimental Biophysics
to explore and quantify molecular recognition and binding 
properties of (bio-)molecular complexes, such as
reaction off-rates, binding energy landscapes,
affinity ranking of single point mutations etc.
\cite{hin06,mer01,rit06}.
Typical examples include ligand-receptor 
compounds like antibody-antigen  \cite{hin96}, 
protein-DNA \cite{eck05}, or supramolecular 
complexes \cite{sch10}.
In all these experiments,
one of the participating
molecules is connected to a tiny force probe 
and the other to a substrate surface 
(or an immobilized micron-sized bead).
By controlling the relative distance $s$ between these objects with nanometer
precision, an experimentally controlled, steadily increasing force $F(s)$ in the 
Piconewton range can be applied to the compound until the chemical 
bonds between the constituting molecules rupture,
see Fig. \ref{fig2}A.

Repeating the same experiment many times results in a 
broad distribution of rupture forces,
whose interpretation and quantitative analysis 
represents a formidable theoretical challenge,
of foremost importance for our understanding of the 
basic biophysical principles at work and for
the practical evaluation of experimental rupture force data.

In other words, given the distribution of rupture forces
for various pulling velocities, a theoretical framework
is needed in order to achieve a quantitative characterization 
of the probed chemical bond in terms of binding affinities, 
force free dissociation rates, and molecular binding 
energy landscape parameters.
The predominant general framework in this context
is due to Evans and Ritchie \cite{eva97}, 
describing the rupture of a molecular complex 
as the dissociation of a {\em single} chemical bond
under the externally applied force in terms of 
a thermally activated escape process with a force 
dependent decay rate.
While Evans and Ritchie adopted the phenomenological 
approach by Bell \cite{bel78},
important subsequent refinements concern e.g.
the specific force dependence of the rate
\cite{hum03,han06,sch06,dud06}, 
multiple dissociation pathways \cite{bar02}, 
linker length distributions \cite{fri03}, 
and contact times \cite{lue06}.

Experimentally, in a given force extension 
curve $F(s)$ it is often quite obvious
that more than one bond has actually been probed (Figs. \ref{fig2}B,C).
Since such multiple rupture events cannot be
quantitatively evaluated within the above
theories, they are usually eliminated as
far as possible from the experimental data 
sets.
The key issue of our present work
is to take into account the possibility that 
even after eliminating multiple bonds along this 
line, the remaining apparent single bonds
may in fact still contain a significant number 
of ``hidden'' multiple bonds.
As we will show, this number is 
indeed far from negligible under usual experimental
conditions.
While a recent experimental work has demonstrated that 
this might indeed be the case \cite{may10}, a thorough
theoretical study of this point has still been
missing so far.
Such a violation of the basic assumption of 
the above standard approach in this
context is not only of conceptual interest, 
but also may have considerable
practical implications regarding the numerous
experimental works employing this 
``standard'' theoretical framework 
for evaluating their data.

Specifically, within this standard 
framework the main focus is usually 
put on the dependence of the most probable 
rupture force $f^\ast$ on the pulling velocity $v$.
While the corresponding predictions indeed 
agree very well with most experiments,
a more careful analysis of the complete rupture force 
distribution reveals that most experimental 
data are actually incompatible with any of those models 
\cite{imp04,rai06}, in particular the typical
``long tails'' \cite{lue06,may10} of the experimental 
rupture force distributions, 
as exemplified by Figs. \ref{fig2}E,F.

A first, purely phenomenological explanation of 
those incompatibilities in terms of bond-heterogeneities 
was proposed in \cite{rai06}, amounting to
a parametric randomization of the force 
dependent single bond dissociation rates.
Our present work provides for the first time 
a more detailed microscopic foundation of this 
ansatz, since a mixture
of ``true'' single bonds and undiscovered multiple 
bonds is similar to a statistical 
ensemble of formal single bonds with 
randomized parameters in the force 
dependent rate (see discussion in Sect. \ref{4.4}).
While hidden multiple bonds thus seem to 
play a prominent role in this context, 
a more detailed analysis, which goes beyond the scope 
of our present paper, however indicates, that there may
well exist additional microscopic mechanisms which 
also contribute notably to the above mentioned 
incompatibilities and their explanation in terms 
of bond-heterogeneities.

Yet another main conclusion of our present paper is
that even though a considerable 
fraction of rupture forces are due to (unresolved) 
multiple bonds, the rate parameters can still be 
extracted from the dependence of the most 
probable rupture force on the pulling 
velocity according to the above mentioned 
single bond theories. In other
words, the hidden multiple bonds 
heavily contribute to the long tails 
in Figs. \ref{fig2} (E) and (F), but hardly to the main peaks.

We should emphasize that
besides the investigations of ligand-receptor 
dissociation, the general realm of so-called 
``dynamic force spectroscopy'' also includes the exploration of
protein unfolding, unzipping of DNA strands 
and DNA hairpins, forced rupture of cell adhesion etc., 
see e.g.  \cite{bel78,erd04,erd07,dud07,dud08} 
and further references therein.
These quite different experimental set ups often 
admit a theoretical treatment quite similar
to those mentioned above.
However, they usually do {\em not} exhibit
any kind of incompatibilities e.g. in the 
form of long tails, see e.g.
\cite{dud07,dud08,sch06,li00,hum03,wil03,koc03,die06}.
Our present approach specifically and exclusively
concerns the case of ``dynamic rupture force spectroscopy'' by AFM,
micropipette-based force probes, 
laser tweezers etc. \cite{mer01,rit06,hin06}.
With the exception of \cite{mor07}, the problem of the long 
tails does arise for all those rupture force experiments 
we know of.

As far as multiple bond models are concerned,
we build on several related previous studies:
A common starting point in many of those works,
see e.g. \cite{bel78,erd04,erd07,sei00,eva01,ger02,wil02,rat06,sul06,erd08,ran08,guo08},
is the assumption that the applied force is 
equally distributed among all bonds.
Closer inspection however shows that 
the distribution of the 
rupture forces, expected from these models, in general exhibits, 
unlike the experimentally observed 
ones (Figs. \ref{fig2}E,F), 
several well separated peaks, which are attributed 
to single, double ,triple, ... bonds \cite{rat06,guo08},
see also Sect. \ref{4.3} below for more details.
In other words, the assumption of equally 
distributed forces must be given up.
Such a refined theoretical treatment has been
developed for the first time by Akhremitchev 
and coworkers \cite{guo08}, but their approach still
exhibits several other strong oversimplifications:
First, the number of receptors still cannot be more 
than one, while the number of ligands is at most two.
Second, in case of two coexisting bonds, both
of them are assumed to rupture simultaneously.
Another important series of works considering
forces which are unequally distributed among several 
coexisting bonds is due to Gao and coworkers
\cite{qia08,che10}.
Formally, they are quite similar to our present
study but physically they focus on cell adhesion
and do not admit any conclusions with respect
to our present case.

Taking into account the very
fact that force extension curves with clearly 
visible multiple bond signatures (Figs. \ref{fig2}B,C)
are discarded before further evaluating those 
data, is yet another essential difference 
between our present work and all the above 
mentioned previous multiple bond studies.

Conceptually, the basic idea of our paper is to 
establish and analyze a simplified but still 
reasonably realistic model of the entire 
experimental procedure:
preparation (functionalization) of the AFM tip
and of the sample surface; formation of bonds;
retraction of the tip from the surface, 
leading to bond dissociation (rupture events);
processing and evaluation of the so obtained 
data exactly as in the case of real 
experimental data.
The advantage of such a ``computer experiment''
is that the interpretation of the data 
and the final conclusions can be immediately
compared with what was actually going on
in the considered system.

%%%%%%%%%%%%%%%%%%%%%%%%%%%%%%%%%%%%%%%%%%%%%%%%%%%%%%%%%%%%%%
\section{Model}
\label{model}
%
%%%%%%%%%%%%%%%%%%%%%%%%%%%%%%%%%%%%%%%%%%%%%%%%%%%%%%%%%%%%
\subsection{Functionalization of tip and sample}
\label{functionalization}
%
%
%%%%%%%%%%%%%%%%%%%%%%%%%%%%%%%%%%%%%%%%%%%%%%%%%%%%%%
%
A typical experimental setup we have in mind is exemplified 
by Fig. \ref{fig1}A. 
As illustrated by Fig. \ref{fig1}B,
the AFM tip is modeled as a half 
sphere of experimentally realistic radius $R=15\,$nm,
to which ligands are connected via
linkers of experimentally realistic
length $L=30\,$nm.
The number of linkers is thus equal to 
the number of ligands and is denoted by $\NL$.
The vector connecting the tip apex with the 
immobilization point of linker $i$ ($i=1\dots \NL$) 
is denoted as $\rL$ and the $\NL$ immobilization points
themselves are randomly sampled according to a
uniform distribution on the half sphere.
In a real experiment, the number $\NL$ of linkers attached 
to the tip is itself a random variable with a distribution 
which depends in a highly complicated way on
the (chemical) functionalization procedure.
Within our model, we assume the linker number 
$\NL$ to be given and we will show that our conclusions 
concerning, e.g., the distribution of rupture forces,
depend very little on the exact value of $\NL$, 
at least as long as $\NL \gtrsim 5$ 
(see Sect. \ref{results}).
In view of typical grafting densities of, e.g., 
the PEG linkers used in \cite{sch10},
a realistic choice which we often adopt
in our numerical calculations below is
$\NL=10$.

The receptors are randomly immobilized 
on the sample surface with no or 
negligibly short linkers.
Beside the case that they are independently and 
uniformly distributed according to some preset
density $\rho_{\mathrm{rec}}$ (Sec. \ref{results}),
we will also discuss the case that the receptors 
are distributed in clusters of varying size 
(Sec. \ref{clusters}).
%
%
%%%%%%%%%%%%%%%%%%%%%%%%%%%%%%%%%%%%%%%%%%%%%%%%%%%%%%%%%%%%%%
\subsection{Force distance cycle}
\label{cycle}
Usually one refers to a single 
repetition/run of a pulling experiment 
as a force distance cycle:
In a first step,
tip and sample are brought into contact
(at the origin of our coordinate system)
for a certain ``dwell-time'' 
$t_{\mathrm{dwell}}$
(typically $0.1 - 1$s),
during which
receptors and ligands may 
form bonds.
Then, the AFM tip is retracted 
along the z-axis,
$\mathbf{r}_{\mathrm{apex}}(t) = (0,\, 0,\, z(t))$
(cf. Fig. \ref{fig1}),
resulting in a steadily increasing force 
on the single bonds until they rupture.
In the force extension curve, 
bond ruptures induce more or less 
pronounced force dips (Figs. \ref{fig2}A-D).
As in the real experiment, if more than one such
dip is clearly visible (Figs. \ref{fig2}B,C), the
corresponding force extension curve is discarded.

Before discussing these steps in more detail 
in the following subsections, we remark that
while the AFM tip and thus the linker immobilization points
$\rL$ remain the same for all cycles (see Fig. \ref{fig1}B),
a new random sampling of the receptor positions
will be performed for each cycle, modeling 
the fact that in a real experiment the sample surface is 
usually ``scanned'' either due to thermal drift or 
controlled lateral displacements.
%
%
%%%%%%%%%%%%%%%%%%%%%%%%%%%%%%%%%%%%%%%%%%%%%%%%%%%%%%%%%%%%%%
\subsection{Formation of bonds}
\label{formation}
We consider $\NL$ ligands, attached by linkers to the AFM tip,
and a surface with randomly distributed receptors,
having the opportunity to form bonds during the dwell 
time $t_{\mathrm{dwell}}$.
The resulting number of bonds thus
depends on the dynamics of a nanometer-sized object 
tethered to a polymer chain, which itself may
interact with other chains, and which is 
subjected to complicated boundary conditions 
(half-sphere on a flat surface) 
and far from equilibrium initial conditions.
Instead of theoretically addressing this very
difficult problem, we invoke the 
experimental findings in a related situation, 
namely the association of end-grafted chains 
to uniformly reactive surfaces
\cite{jep01,abb06,guo09}.
As a result of these works, the probability that a 
bond between a tethered molecule and one of its 
potential binding partners at a distance $d$ is 
actually formed is 
approximately constant for $d<d_{\mathrm{max}}$
and negligible for $d>d_{\mathrm{max}}$, where
$d_{\mathrm{max}}\approx$ 10-15\,nm
for PEG linkers with $L \approx 30\,$nm and  
$t_{\mathrm{dwell}} \approx 1$s.
Moreover, the number of remaining ``unpaired potential 
binding partners'' closer than $d_{\mathrm{max}}$ 
is relatively small.

We therefore proceed as follows:
First, all ligand-receptor pairs with a distance 
smaller than $d_{\mathrm{max}}$ are identified.
Then, one of these pairs is randomly chosen 
to form a bond.
For the remaining receptors and ligands these two 
steps are repeated,
until no pair with a distance smaller than 
$d_{\mathrm{max}}$ is left.
The total number of the so obtained bonds is henceforth
denoted as $N$.
In our numerical calculation below, 
our standard choice will be
$d_{\mathrm{max}} = 12\,\mbox{nm}$,
while a more detailed exploration of other
choices will be discussed in Sect. \ref{4.3}.

We have also tried out other strategies for the 
bond formation, for example that a ligand 
preferentially forms a bond with the closest 
receptor. 
This, however, yielded very similar results to the 
above mentioned procedure.

%%%%%%%%%%%%%%%%%%%%%%%%%%%%%%%%%%%%%%%%%%%%%%%%%%%%%%%%%%%%%%
\subsection{Force extension curves}
\label{forces}
Denoting by $\rL$ and $\rR$ the linker and receptor 
immobilization points of the 
$i$-th bond ($i=1...N$), the linker's 
end-to-end vector is (see Fig. \ref{fig1}B)
\begin{eqnarray}
  \label{10}
  \mathbf{r}_{i} &=& z\mathbf{e}_z
  + \rL 
  - \rR  =  l_i \, \mathbf{\hat{r}}_{i} \ ,
\end{eqnarray}
where $\mathbf{e}_z$ is the unit vector in $z$-direction,
$\mathbf{\hat{r}}_{i}$ is the unit vector in $ \mathbf{r}_{i}$-direction,
and $l_i$ is the length (modulus) of $\mathbf{r}_{i}(z)$.

Due to entropic effects, stretching the linker to a length $l_i$
requires a force of modulus $f_i$ which acts in the direction
of the end-to-end vector, i.e. 
\begin{equation}
\mathbf{f}_i = f_i\mathbf{\hat{r}}_{i} \ .
\label{11b}
\end{equation}
A particularly simple model for this force-dependence of the linker 
length is the freely jointed chain (FJC) model, according to which
 \begin{equation}
  \label{12}
  l_i(f_i) = L \left[
    \coth\left(\frac{\lambda_Kf_i}{k_BT}\right) -\frac{k_BT}{\lambda_K f_i}
  \right] \ ,
\end{equation}
where $\lambda_K$ is the Kuhn length and $k_BT$ the thermal energy.
As a typical value, obtained
by stretching experiments on PEG linkers 
at room temperature \cite{mer01}, we  
employ $k_BT/\lambda_K = 6\,\mbox{pN}$.
Numerically inverting Eq. \ref{12} then
yields the force extension characteristic 
$f_i(l_i)$ of a single linker which we henceforth 
will utilize in our model.

For the pulling geometries studied in our present work, 
the experimentally accessible observable is not
the magnitude of the total force
$\mathbf{F} = \sum_{i=1}^N  \mathbf{f}_i$,
but rather it's normal component with respect
to the sample surface \cite{kuh06},
i.e.
\begin{equation}
  \label{13}
  F := \mathbf{F}\cdot \mathbf{e}_z  = 
\sum_{i=1}^N (\mathbf{\hat{r}}_{i} \cdot \mathbf{e}_z)\, f_i \ .
\end{equation}
A priori, all quantities in Eqs. \ref{10}-\ref{13}
are functions of $z$.
However, the actual control parameter in the 
experiment is the displacement 
\begin{equation}
s = vt
\label{13a}
\end{equation}
of the sample surface from its initial position,
with a typical experimental pulling  velocity
$v = 1000\, \mbox{nm/s}$.
Furthermore, the displacement $s$ can be written as
the sum of $z$ and the deflection of the cantilever 
under the force $F$, i.e.
\begin{equation}
  \label{14}
  s = z + F/\kappa\ ,
\end{equation}
with a typical experimental cantilever stiffness \cite{mer01}
of $\kappa = 10\,\mbox{pN/nm}$.
Via Eq. \ref{14}, all the above quantities, in particular 
the force $F$ from Eq. \ref{13}, may thus alternatively be
expressed as functions of $s$, and likewise, 
via Eq. \ref{13a}, as functions of $t$.
In the following, we will switch between these 
alternatives without much further ado.
In particular, $F(s)$, following from Eq. \ref{13} with
$s$ as independent variable, is called the force 
extension curve.

%
%
%%%%%%%%%%%%%%%%%%%%%%%%%%%%%%%%%%%%%%%%%%%%%%%%%%%%%%%%%%%%%%
\subsection{Rupture probabilities}
\label{single_bond}
For an arbitrary but fixed bond $i$,
the force $f_i$ appearing in Eq. \ref{13} follows by 
inverting \ref{12} before the bond has ruptured
and is given by $f_i=0$ after rupture.
According to Evans and Ritchie \cite{eva97},
the rupture process itself is a thermally activated 
rate process, governed by
\begin{equation}
  \label{1}
  \dot{n}_i(t) = -k(f_i(t)) n_i(t), \hspace*{0.3cm}
  n_i(t=0) = 1\ ,
\end{equation}
where $n_i(t)$ denotes the survival probability of the bond
up to time $t$, and $k(f_i(t))$ is the force dependent 
bond-dissociation rate.
Consequently,
\begin{equation}
  \label{2}
  n_i (t) = \exp\left[
  -\int_0^t dt'\; k(f_i(t'))
  \right] 
\end{equation}
and the probability of bond rupture then follows as $-d\,n_i(t)/dt$.
Denoting by $t_i(f)$ the inverse of the function $f_i(t)$, 
one furthermore obtains the survival probability 
$n_i(f):=n_i(t_i(f))$ of the $i$-th bond
as a function of the instantaneous
force $f$ acting on that bond. 

In our quantitative calculations below,
we will adopt the simplest and most common 
approximation \cite{bel78}
\begin{equation}
  \label{3}
  k(f) = k_0 \exp(x_b f / k_BT) \ ,
\end{equation}
where $k_0$ is the force-free dissociation rate and
$x_b$ the distance between potential minimum and 
barrier along the direction of the applied force.
Furthermore, we will use the
following typical values of those rate parameters:
\begin{equation}
k_0=0.1\,\mbox{s}^{-1}\ , \ \ x_b/k_BT=0.1\,\mbox{pN}^{-1} \ .
\label{3a}
\end{equation}
We remark that the alternative approximations for $k(f)$ from 
\cite{dud06} did not lead to notably different findings than Eq. \ref{3}.
%
%
%
%%%%%%%%%%%%%%%%%%%%%%%%%%%%%%%%%%%%%%%%%%%%%%%%%%%%%%%%%%%%%%
\subsection{Simulation and processing of force extension curves}
\label{processing}
The initial condition in the form of $N$ bonds 
at time $t=0$ is set according to Sect. \ref{formation}.
For the further time evolution we adopt --
similarly as in \cite{erd04,erd07,qia08,che10} -- 
Gillespie's algorithm \cite{gil77} with time-dependent forces
and rupture probabilities as detailed in Sects. 
\ref{forces} and \ref{single_bond}, respectively.
%%%%%%%%%%%%%%%%%%%%%%%%%%%%%%%%%%%%%%%%%%%%%%%%%%%%%%%%%%%%%%

Fig. \ref{fig5} exemplifies six force extension 
curves simulated along these lines.
They indeed look quite similar to the 
experimental curves from Figs. \ref{fig2}A-D.
In particular, while in the left panels of 
Fig. \ref{fig5} the two rupture 
events (``dips'' or ``jumps'') happen to be well separated, 
the two bonds accidentally rupture nearly simultaneously in the right panels.
In the middle panels, the rupture of the first bond happens to be still 
visible as a small force dip, but due to the limited resolution of a real
experimental device (cf. Fig. \ref{fig2}D), 
it is questionable whether in practice such a 
force signal could be distinguished from that of a single bond.

For comparison, the bottom right panel in Fig. \ref{fig5} also shows a 
typical force extension curve from a ``numerical experiment'' 
with a single bond ($N=1$).
This example corresponds to the more likely situation that 
the single bond ruptures earlier than the two bonds, but 
since these are random events with a quite
broad distribution (see later sections), 
it might also be the other way round with 
quite some probability.
Within the experimentally realistic noise level
which has been used for the illustration in
Fig. \ref{fig5}, it thus is indeed
practically impossible to
tell on the basis of the dotted and the solid lines
in the right and middle panels of Fig. \ref{fig5}, 
which one was originally due to a double and a single 
bond, respectively. 

Similar observations apply to more than two bonds.
In other words, it seems reasonable to assume that the rupture of
a multiple bond complex cannot be distinguished from that of a single bond
if all bonds rupture within some small 
distance $\Delta s_{\mathrm{max}}$.
In view of Figs. \ref{fig2}A-D and \ref{fig5},
$\Delta s_{\mathrm{max}} = 1\,\mbox{nm}$
seems to be an optimistic but still reasonable choice for the resolution limit,
in good agreement with the experimental value of
$\Delta s_{\mathrm{max}} \approx 4\,$nm reported in 
\cite{may10} for relatively long linkers.
A more detailed discussion of the specific choice of
$\Delta s_{\mathrm{max}}$ will be provided in 
Sect. \ref{4.3}.

A further consequence of the instrumental noise is that
only rupture forces beyond some threshold value
$f_{\mathrm{min}}$ can be clearly distinguished from the noise.
In our numerical examples below, we will always 
adopt the experimentally realistic choice
$f_{\mathrm{min}}=20\,\mbox{pN}$ since we
found that any other choice of
$f_{\mathrm{min}}$ within reasonable 
limits (say between 0\,pN and 40\,pN)
hardly affected the results with the obvious
exception of different ``lower cutoffs'' 
in rupture force histograms like Figs. \ref{fig2}E,F.

As in real experiments, force extension curves
which exhibit clearly resolvable signatures of 
multiple bonds are excluded from the further analysis.
In turn, this means that force extension curves
of multiple bonds will still be accepted
if all bonds with rupture force larger than 
$f_{\mathrm{min}}$ rupture within a distance 
$\Delta s_{\mathrm{max}}$.
Such multiple bonds, which are erroneously
classified as single bonds will henceforth 
be denoted as ``false single bonds'' or ``hidden multiple bonds''.
In turn, both the ``true'' and ``false''
single bonds will be called ``apparent single bonds''.
The force value, at which an apparent single bond
ruptures is defined by the maximum of the 
force extension curve $F(s)$,
is termed ``rupture force'',
and is conventionally denoted by the letter $f$.
Due to the random features of tip functionalization
(Sect. \ref{functionalization}), bond formation 
(Sect. \ref{formation}), and bond rupture
(Sects. \ref{single_bond} and \ref{processing}),
the rupture force $f$ is a random variable,
whose probability density is
henceforth denoted as $p(f)$.

%%%%%%%%%%%%%%%%%%%%%%%%%%%%%%%%%%%%%%%%%%%%%%%%%%%%%%%%%%%%%%
\section{Uniformly distributed receptors}
\label{results}
%%%%%%%%%%%%%%%%%%%%%%%%%%%%%%%%%%%%%%%%%%%%%%%%%%%%%%%%%%%%%%
%
For a uniform receptor density $\rho_{\mathrm{rec}}$
(cf. Sect. \ref{functionalization}), the 
number of receptors on a surface area $A$ 
is a Poisson distributed random number 
with mean value $\rho_{\mathrm{rec}} A$.
Sampling receptors along these lines, 
we simulated a number of force distance 
cycles so that 1000 of them were exhibiting at 
least one rupture event (i.e. $N\geq 1$).
In Figs. \ref{fig6}A,B we moreover averaged
over 100 different AFM tips (cf. Sect. \ref{functionalization}),
while Figs. \ref{fig6}C,D depict representative
results for one single tip.

The blue line in Fig. \ref{fig6}A shows that the 
probability of classifying an actual multiple bond 
as an apparent single bond is 40-50\% and decreases 
only weakly with increasing density $\rho_{\mathrm{rec}}$ of receptors.
Likewise, the fraction of ``hidden multiple bonds''
among all apparent single bonds (black line in Fig. \ref{fig6}A)
increases remarkably fast as a function of $\rho_{\mathrm{rec}}$.
However, for these $\rho_{\mathrm{rec}}$ values, also the probability
to observe at least one rupture event during one force 
distance cycle (red line in Fig. \ref{fig6}A) is
quite large compared to the usual experimental findings.
We come back to this point in Sec. \ref{4.1}.

%%%%%%%%%%%%%%%%%%%%%%%%%%%%%%%%%%%%%%%%%%%%%%%%%%%%%%%%%%%%%%

Fig. \ref{fig6}B shows that the rupture force, averaged 
over all apparent single bonds, exhibits a moderately
increasing behavior as a function of 
the receptor density $\rho_{\mathrm{rec}}$.
This finding is quite plausible in view of the increasing 
number of hidden multiple bonds, which are expected
to rupture on the average at higher force values than the 
true single bonds.

For small receptor densities $\rho_{\mathrm{rec}}$, almost only 
true single bonds contribute to the distribution 
of rupture forces $p(f)$
which is consequently sharply peaked and exhibits a fast
decay for large $f$ (Fig. \ref{fig6}C).
For higher receptor densities, the distribution of rupture forces 
develops a long tail (Fig. \ref{fig6}D)
due to the increasing fraction of hidden multiple bonds,
much like for the real data exemplified by Figs. \ref{fig2}E,F.
Interestingly, the distribution still exhibits a pronounced peak
which (by comparison to Fig. \ref{fig6}C) 
is mainly caused by the rupture events of the true single bonds,
while no distinct additional peaks due to double and more general
multiple bonds are visible. 

While Fig. \ref{fig6} was obtained for AFM tips with $\NL=10$ 
linkers,
Fig. \ref{fig7} shows that 
$\NL=5$ and $\NL=15$ lead only to
minor changes of these results.

%%%%%%%%%%%%%%%%%%%%%%%%%%%%%%%%%%%%%%%%%%%%%%%%%%%%%%%%%%%%%%
\section{Clustering of receptors}
\label{clusters}
%%%%%%%%%%%%%%%%%%%%%%%%%%%%%%%%%%%%%%%%%%%%%%%%%%%%%%%%%%%%%%
\subsection{Motivation}
\label{4.1}
%%%%%%%%%%%%%%%%%%%%%%%%%%%%%%%%%%%%%%%%%%%%%%%%%%%%%%%%%%%%%%
Experimentally, the receptor density 
is controlled via the chemical preparation procedure 
of the sample surface. 
Usually it is adjusted such that the probability 
to observe at least one rupture event 
during one force distance cycle 
(red lines in Figs. \ref{fig6} and \ref{fig7})
is quite low (typically below 20\%).
Assuming a uniform distribution of receptors 
on the surface, the probability of multiple 
bonds would then be negligibly small, cf. 
Figs. \ref{fig6} and \ref{fig7}.
It can be shown that this observation is largely independent
of the remaining model parameters and does also
apply to other geometries of the AFM tip.
Yet, in real experiments 
multiple bonds
are actually observed at a much higher rate.
This seems to us a quite convincing direct evidence 
that the receptors are in fact
not uniformly distributed on the surface.
Physical reasons of why the density of receptors 
may be {\em locally} considerably larger 
than on the average are:
(i) The receptors (or the linkers used to immobilize them on 
the substrate) may tend to cluster due to mutual 
interactions, insufficient mixing 
or substrate inhomogeneities (some areas may
be more ``reactive'' than others).
(ii) One receptor may exhibit several identical binding sites for 
the ligand, like e.g., streptavidin for biotin \cite{guo08}.

%%%%%%%%%%%%%%%%%%%%%%%%%%%%%%%%%%%%%%%%%%%%%%%%%%%%%%%%%%%%%%
\subsection{Modeling}
\label{4.2}
%%%%%%%%%%%%%%%%%%%%%%%%%%%%%%%%%%%%%%%%%%%%%%%%%%%%%%%%%%%%%%
Receptor clusters are uniformly distributed
on the sample surface according to some cluster 
density $\rho_{\mathrm{clu}}$.
For any given cluster, the number of receptors 
$\NC$, is sampled from a Poisson distribution
with mean $\langle\NC\rangle$ and the receptors 
are then independently distributed within a 
circle of radius
$\RC=2\,\mbox{nm}$.
Other values of $\RC$ lead to very similar 
results, as long as they remain smaller than
the maximal binding length $d_{\mathrm{max}}$ 
(cf. Sect. \ref{formation}).
Likewise, we will always focus on the specific 
cluster density 
$\rho_{\mathrm{clu}}=10^{-4}\,\mbox{nm}^{-2}$
since the value of $\rho_{\mathrm{clu}}$ is in fact irrelevant
as long as multiple bonds involving receptors from different 
clusters are negligible.
Finally, we restrict ourselves to average cluster sizes 
$\langle \NC \rangle = 2$
and discuss the influence of the parameters
$d_{\mathrm{max}}$ and $\Delta s_{\mathrm{max}}$ 
in some more detail.

%%%%%%%%%%%%%%%%%%%%%%%%%%%%%%%%%%%%%%%%%%%%%%%%%%%%%%%%%%%%%%
\subsection{Results}
\label{4.3}
%%%%%%%%%%%%%%%%%%%%%%%%%%%%%%%%%%%%%%%%%%%%%%%%%%%%%%%%%%%%%%

Fig. \ref{fig8} shows numerical results
for different values of the maximal binding length
$d_{ \mathrm{max}}$.
Proceeding as detailed in Sects. \ref{model}
and {\ref{4.2}, we simulated a number of force distance 
cycles so that 10000 of them were exhibiting at 
least one rupture event (i.e. $N\geq 1$), but
unlike in Sect. \ref{results}, no additional
sampling over different AFM tips was performed.
The resulting probabilities of observing at least one
rupture event during a force distance cycle 
were approximately 5\% for $d_{ \mathrm{max}}=8\,$nm,
11\% for $d_{ \mathrm{max}}=12\,$nm,
and 17\% for $d_{ \mathrm{max}}=16\,$nm.
As mentioned in Sect. \ref{4.1}, these are typical values 
for real experiments.

Since resolution limits $\Delta s_{ \mathrm{max}} \lesssim 1\,$nm 
seem to be experimentally unrealistic (cf. Figs. \ref{fig2}A-D, \ref{fig5}),
Fig. \ref{fig8}A implies that a large fraction
of actual multiple bonds will be classified as apparent 
single bonds, and Fig. \ref{fig8}B shows that
a considerable fraction of apparent single bonds
are actually multiple bonds.

Also the rupture force distributions $p(f)$
(Fig. \ref{fig8}C) seem to depend only 
quite weakly on the maximal binding length 
$d_{ \mathrm{max}}$, apart from two details: 
(i) In contrast to Fig. \ref{fig6}D,
the rupture force distributions develop 
a small secondary peak upon decreasing 
$d_{ \mathrm{max}}$.
The reason is that for small $d_{ \mathrm{max}}$,
the force is always more or less equally distributed 
among the multiple bonds, giving rise to well 
separated peaks for single, double, triple, ... bonds.
With increasing $d_{ \mathrm{max}}$ values, 
multiple bonds with unequally distributed forces
become more likely, hence the force peaks are smeared 
out and finally disappear.
In either case, 
it is questionable whether the minima of the
rupture force distributions $p(f)$
in Fig. \ref{fig8}C can also be resolved in 
practice, where sample sizes are in general much smaller 
(see also Figs. \ref{fig2}E,F and \ref{fig9}A-D) and where
the peaks are broadened by noise.
(ii) The complete rupture force distribution $p(f)$ 
for $d_{ \mathrm{max}}=16\,$nm is slightly shifted 
towards lower forces compared to the distribution
for $d_{ \mathrm{max}}=8\,$nm.
The reason is that the rupture force corresponds to the 
force measured by the AFM cantilever and not to the
force acting on the bond, see Fig. \ref{fig1}B.
For larger values of $d_{ \mathrm{max}}$ it is more probable 
that one is pulling under a large angle 
relative to the surface normal (see Fig. \ref{fig1}B),
resulting in smaller values of the measured rupture
forces.

%%%%%%%%%%%%%%%%%%%%%%%%%%%%%%%%%%%%%%%%%%%%%%%%%%%%%%%%%%%%%%
\subsection{Variations of the pulling velocity}
\label{4.4}
%%%%%%%%%%%%%%%%%%%%%%%%%%%%%%%%%%%%%%%%%%%%%%%%%%%%%%%%%%%%%%

Fig. \ref{fig9} shows the numerical imitation of a complete
dynamic force spectroscopy experiment:
For each pulling velocity $v$ 
we simulated 500 force distance cycles 
as detailed in the preceding sections,
employing $d_{max}=12\,$nm, $\Delta s_{\mathrm{max}}=1\,$nm, 
and the same AFM tip as in Fig. \ref{fig8}.
The resulting probabilities of observing at least one
rupture event during a force distance cycle 
were approximately 10\% for all pulling velocities $v$.

While a detailed quantitative comparison/fitting 
with any specific experiment is not the purpose of 
our present paper, the general similarity between
the typical experimental data from Figs. \ref{fig2}E,F
and the numerical simulations in Figs. \ref{fig9}A-D
is quite convincing.
In particular, we recover the typical ``long tails''
announced in Sect. \ref{introduction}.

For a given pulling velocity $v$,
the survival probability $n(f)$ of the apparent (``true'' or ``false'') 
single bonds readily follows from the rupture force distribution
$p(f)$ according to 
\begin{equation}
n(f)=\int_f^\infty p(f')\, df' \ .
\label{20}
\end{equation}
Fig. \ref{fig9}E presents those survival probabilities
for various pulling velocities $v$.
As demonstrated in \cite{rai06}, 
all these functions
$-v\ln p(f)$ must collapse onto a single, $v$-independent 
``master curve'' in the absence of false single bonds,
while experimentally they actually split in a very
similar way to the curves shown in 
Fig. \ref{fig9}E. 
In \cite{rai06} this was explained by adopting a parametric randomization 
of the dissociation rate from Eq. \ref{3}.
In our case, the dissociation rate for a single bond $i$ can be written as 
$k_0\exp[ (x_ba_i)F/k_BT]$ where $a_i:=f_i/F$. 
For a fixed force $F$, $a_i$ is a random variable 
depending on the number of parallel bonds and their configuration.
Effectively, hidden multiple bonds thus result in very
similar rupture force distributions as bond heterogeneities. 

The fact that the curves in Fig. \ref{fig9}E indeed do collapse
onto a single master curve for small $f$-values further corroborates
the above discussed implications of Figs. \ref{fig6} C and D, namely
that the hidden multiple bonds mainly affect the
rupture force distribution beyond its maximum, while
the maximum itself is mainly governed by the ``true'' 
single bonds.

%%%%%%%%%%%%%%%%%%%%%%%%%%%%%%%%%%%%%%%%%%%%%%%%%%%%%%%%%%%%%%
\subsection{Most probable rupture forces}
\label{4.5}
%%%%%%%%%%%%%%%%%%%%%%%%%%%%%%%%%%%%%%%%%%%%%%%%%%%%%%%%%%%%%%
As seen above, hidden multiple bonds 
hardly affect the maximum of the rupture 
force distribution.
In other words, the most probable rupture 
force $f^\ast$ admits an adequate and 
consistent theoretical treatment within
the ``traditional''  single bond picture
by Evans and Ritchie \cite{eva97}, 
cf. Sect. \ref{introduction}.
In particular, the rate parameters 
$x_b/k_BT$ and $k_0$ appearing in Eq. \ref{3a} 
can still be estimated by plotting
$f^\ast$ versus $\ln \lambda$
due to the well established relation
\begin{equation}
f^\ast = (k_BT/x_b)\, \ln\left(x_b \lambda/k_BTk_0\right)
\label{21}
\end{equation}
where $\lambda := F'(s(f^\ast)) v$ is the so-called loading 
rate \cite{eva97}.

Fig. \ref{fig9}F confirms that the most probable 
rupture force resulting from our simulations indeed
depends linearly on $\ln \lambda$.
By fitting a straight line through these points
one recovers by means of Eq. \ref{21} the following
estimates 
\begin{equation}
k_0 = 0.17\,\mbox{s}^{-1} \ , \ \ x_b/k_B T = 0.0989\,\mbox{pN}^{-1} \ ,
\label{22}
\end{equation}
in good agreement with the original, ``true'' rate parameters
from Eq. \ref{3a}.

%%%%%%%%%%%%%%%%%%%%%%%%%%%%%%%%%%%%%%%%%%%%%%%%%%%%%%%%%%%%%%
\section{Discussion}
\label{discussion}

By a detailed modeling and simulation of a complete dynamic
force spectroscopy experiment -- including the formation of bonds, their 
dissociation under load, and the post processing of the force extension curves --
we have shown that multiple bonds cannot be detected with sufficient
reliability on the basis of the experimentally accessible information,
namely force extension curves exhibiting several distinct force dips
(Figs. \ref{fig2}A-D, \ref{fig5}).
In particular, in order to explain the typical experimentally observed 
frequencies of force distance cycles exhibiting zero, one, 
and more than one rupture events, we found that assuming 
some kind of receptor and/or linker clustering seems
unavoidable (Sect. \ref{4.1}).
As a consequence, a quite reliable indicator that 
a significant number of multiple bonds are 
misinterpreted as single bonds is
a non-negligible fraction of force 
extension curves with experimentally
resolvable multiple dips.

For the sake of simplicity, we have assumed that the receptors are immobilized 
directly on the substrate and that the ligands are 
connected to the AFM tip via linkers of fixed length.
Under typical experimental conditions, 
about half of the multiple rupture events 
then turned out to be undetectable, largely independent 
of the association dynamics and other details of the modeling.
Cases when both, receptors and ligands are 
immobilized via linkers, which may furthermore
exhibit a broad length distribution, 
are considerably 
more difficult to model \cite{bub09},
but intuitively we still expect comparable 
fractions of hidden multiple bonds, in
good agreement with recent experimental 
observations \cite{may10}.

A somewhat related problem in the context of protein unfolding
has recently been addressed by Dietz and Rief \cite{die07}.
These authors have connected the protein of interest to a
modular protein chain resulting in characteristic sawtooth patterns
in the force extension curves.
Employing an objective pattern recognition algorithm, 
Dietz and Rief then used these ``fingerprints''
to identify the ``true'' single molecule unfolding events. 
Similar experimental and theoretical attempts 
in case of dynamic force spectroscopy are not known to the present authors.
But at least 
we can provide a way to ``live'' with those
apparently unavoidable ``hidden'' multiple bonds, namely
by focusing on the most probable rupture force and
disregarding the ``long tails'' of the full rupture 
force distribution.
In other words, our work provides a solid justification 
of what many experimentalists have always been doing 
anyway.
Finally, Zhu and coworkers \cite{che98}
have pointed out already in 1998 
that it is possible to discern single from multiple 
bonds in micropipette measurements just by visually 
monitoring the detachment of the cell.
They further showed, that by varying the waiting time 
for bond formation, it is possible to
extract the kinetic rates and the mean number of bonds. 
This however required a precise control of the density and 
distribution of receptors and ligands, as well as a correct 
stochastic description of the association process. 
Analogous options in the case of measurements by AFM,
where this is not available (see Sects. \ref{formation} and \ref{4.1}),
are not known to us.

\vspace*{0.3cm}

\centerline{\bf Acknowledgments}
\vspace*{0.3cm}
Special thanks is due to D. Anselmetti and A. Bieker
for providing the data presented in Fig. \ref{fig2}.
We also thank A. Fuhrmann, R. Ros, and V. Walhorn for helpful 
discussion.
Financial support by the Deutsche Forschungsgemeinschaft within the 
Collaborative Research Center SFB 613 is gratefully acknowledged.

%%%%%%%%%%%%%%%%%%%%%%%%%%%%%%%%%%%%%%%%%%%%%%%%%%%%%%%%%%%%%%%%%%%%%%%%%%%%%

%%%%%%%%%%%%%%%%%%%%%%%%%%%%%%%%%%%%%%%%%%%%%%%%%%%%%%%%%%%%%%
\newpage
{\large \bf Figure Legends:}\\[0.1cm]

{\bf Figure \ref{fig2}:}
\\
(A)-(D) Typical force extension curves 
from four ``single runs'' of a dynamic 
force spectroscopy experiment by AFM. 
Ideally, the force $F(s)$ steadily grows with 
increasing distance $s$ until the chemical 
bond ruptures (A).
In (B) and (C) more than one force dip
(``downward jump'' of $F(s)$) is clearly 
visible, indicating 
that more than a single bond was involved.
Curves like in (D) would commonly still be
attributed to a single bond rupture 
within the given noise level and 
resolution limit.
The depicted experimental data in (A)-(D) for 
a protein-DNA complex (PhoB mutant and
target sequence, 
pulling velocity of $2000\,$nm/s, 
cantilever stiffness $13\,$pN/nm,
linker length $30\,$nm)
have been kindly provided by A. Bieker 
and D. Anselmetti (Bielefeld University).
(E) and (F) 
Typical histograms of experimental rupture force distributions 
for two different pulling velocities $v$ \cite{eck05}.
After eliminating all the experimental force extension curves
with clearly visible multiple bond signatures as those in (B) and (C), 
the number of remaining ``apparent single bond 
rupture events'' contributing to (E) was 202, and 151 for (F).
The main features are a pronounce first peak 
(most probable rupture force),
vague indications of possible secondary 
peaks, and a ``long tail''.
\vspace*{0.6cm} 

{\bf Figure \ref{fig1}:}
\\
(A) Schematic sketch of dynamic force 
spectroscopy by AFM: a chemical bond of
interest, e.g. in a ligand-receptor complex, 
is connected via two linker molecules with 
the tip of an AFM cantilever and a piezoelectric 
element at distance $s$.
The latter is employed for ``pulling down'' the attached 
linker molecule at some constant velocity $v$ 
which in turn leads to an elastic reaction
force $F(s)$ of the cantilever,
determined from the deflection of a laser beam.
(B) Illustration of the model for multiple parallel bonds. 
The AFM tip is modeled as a half sphere and
forces $\mathbf{f}_i$ act on the ligand-receptor bonds.
\vspace*{0.6cm}

\newpage

{\bf Figure \ref{fig5}:}\\
The dashed lines exemplify
six realizations of force extension curves 
for double bonds ($N=2$), numerically simulated
as described in Sect. \ref{model}. 
Receptors were uniformly distributed on the sample
surface with density $\rho_{\mathrm{rec}}=10^{-3}\,$nm$^{-2}$.
For a better comparability with the experimental curves
from Figs. \ref{fig2}A-D,
we sampled the force extension curves in regular time steps of
$\Delta t =0.1$ms and added a Gaussian (thermal)
noise with standard 
deviation $\sigma_f=\sqrt{\kappa k_B T}= 6.4$ pN \cite{mer01}. 
After that, a running average over 0.5ms was calculated,
imitating the effect of an experimental low pass filter, and
resulting in the solid lines.
The distance $\Delta s$ between the rupture of the two bonds
is indicated in each figure. 
The dotted line in the bottom right panel exemplifies one
realization of a force extension curve for a single 
bond ($N=1$, cf. Fig. \ref{fig2}A).
The somewhat larger fluctuations observed in Figs. 
\ref{fig2}A-D can be attributed to instrumental noise
on top of the thermal noise.
\vspace*{0.6cm} 

{\bf Figure \ref{fig6}:}\\
(A) Red: Probability of formation (and rupture) of at least 
one bond (i.e. $N\geq 1$) within one force distance cycle.
Black: Probability of observing a ``false single bond''
(i.e. an apparent single bond is {\em de facto} a (hidden) multiple bond).
Blue: Fraction of ``false single bonds'' 
among all multiple bonds.
All three probabilities are presented for
various values of the density $\rho_{\mathrm{rec}}$
of uniformly distributed receptors
and have been obtained as detailed in Sects. \ref{model} and \ref{results}.
The error bars indicate the statistical spread 
(standard deviation) due to
our sampling of 100 different tips (see main text).
(B) The corresponding mean rupture forces $\langle f\rangle$.
(C) Representative rupture force distribution for
one AFM tip (see main text) and
$\rho_{\mathrm{rec}} = 2\cdot 10^{-4}\,$nm$^{-2}$.
(D) Same for $\rho_{\mathrm{rec}} = 10^{-3}\,$nm$^{-2}$.
\vspace*{1cm}

{\bf Figure \ref{fig7}:}\\
Same as Figs. \ref{fig6} (A) and (B), but 
for $\NL=5$ linkers in (A) and (B), and
$\NL=15$ linkers in (C) and (D).
\vspace*{1cm} 
\newpage

{\bf Figure \ref{fig8}:}\\
(A) Fraction of ``false single bonds'' 
among all multiple bonds
versus resolution limit
$\Delta s_{\mathrm{max}}$ for maximum binding lengths
$d_{ \mathrm{max}}=8\,$nm (solid), 
$d_{ \mathrm{max}}=12\,$nm (dotted),
and $d_{ \mathrm{max}}=16\,$nm (dashed).
For further details regarding the employed receptor clustering 
model see main text.
(B) Probability of observing a ``false single bond''
(i.e. an apparent single bond is {\em de facto} a (hidden) multiple bond).
(C) Rupture force distributions for 
$\Delta s_{\mathrm{max}}=1\,$nm. 
For reasons of better visibility, the distribution 
for $d_{ \mathrm{max}}=12\,$nm is not shown.
\vspace*{1cm} 

{\bf Figure \ref{fig9}:}\\
(A)-(D) Rupture force distributions for different pulling velocities $v$,
assuming clustering of receptors.
For further simulational details see main text.
(E) The corresponding survival probabilities according to Eq. \ref{20}.
Velocities increase in the direction indicated by the arrow.
(F) The most probable rupture force $f^\ast$ from (A)-(D) 
versus logarithm of the loading rate $\lambda$. 
The solid line shows the best linear fit.
For more details see main text.
%%%%%%%%%%%%%%%%%%%%%%%%%%%%%%%%%%%%%%%%%%%%%%%%%%%%%%%%%%%%%%

\clearpage

\begin{figure}
\epsfbox{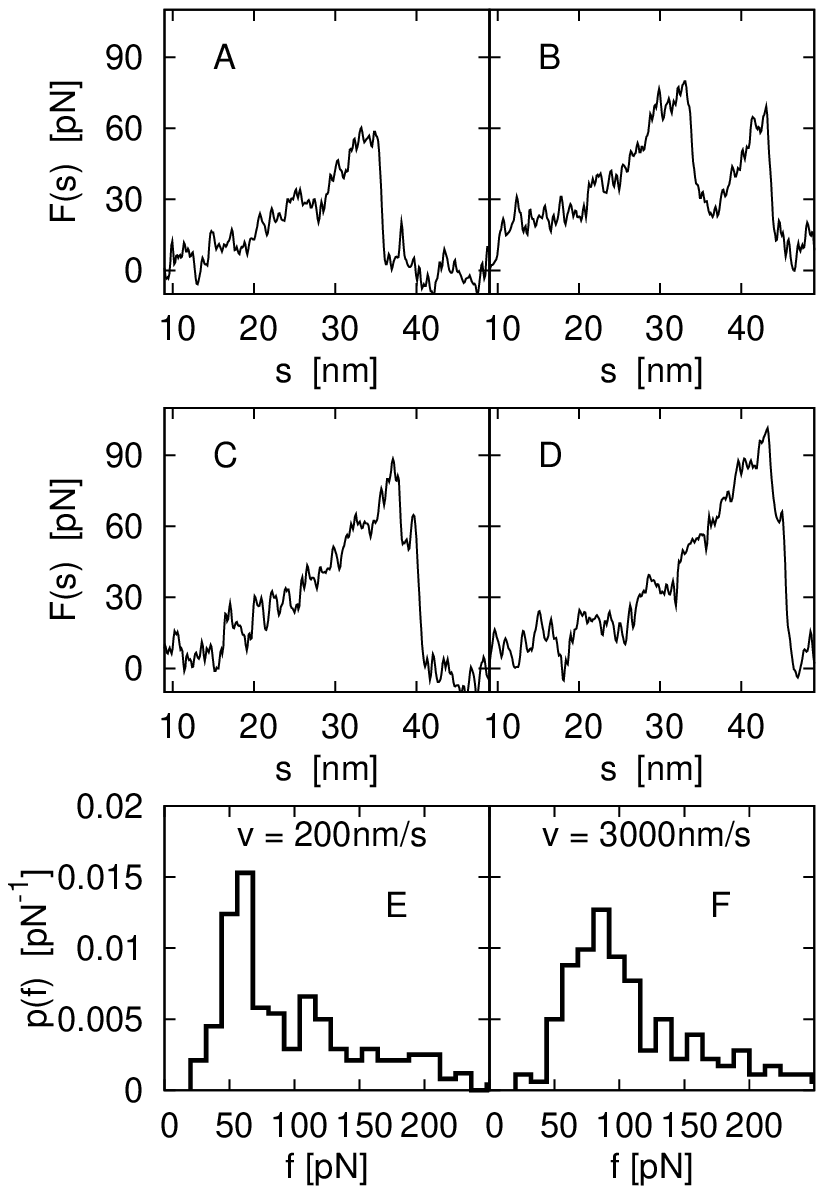}
\caption{
}
\label{fig2}
\end{figure}
%%%%%%%%%%%%%%%%%%%%%%%%%%%%%%%%%%%%%%%%%%%%%%%%%%%%%%%%%%%%%%
\begin{figure}
\epsfbox{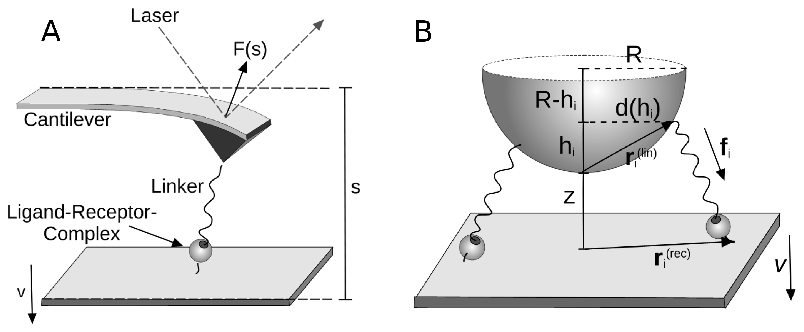}
\caption{
}
\label{fig1}
\end{figure}
%%%%%%%%%%%%%%%%%%%%%%%%%%%%%%%%%%%%%%%%%%%%%%%%%%%%%%%%%%%%%%
\begin{figure}
\epsfbox{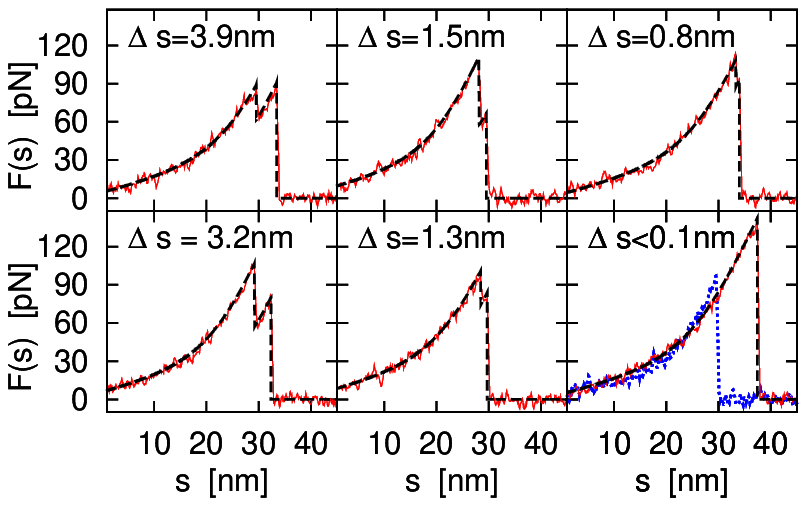} 
\caption{
}
\label{fig5}
\end{figure}
%%%%%%%%%%%%%%%%%%%%%%%%%%%%%%%%%%%%%%%%%%%%%%%%%%%%%%%%%%%%%%
\begin{figure}[t]
\epsfbox{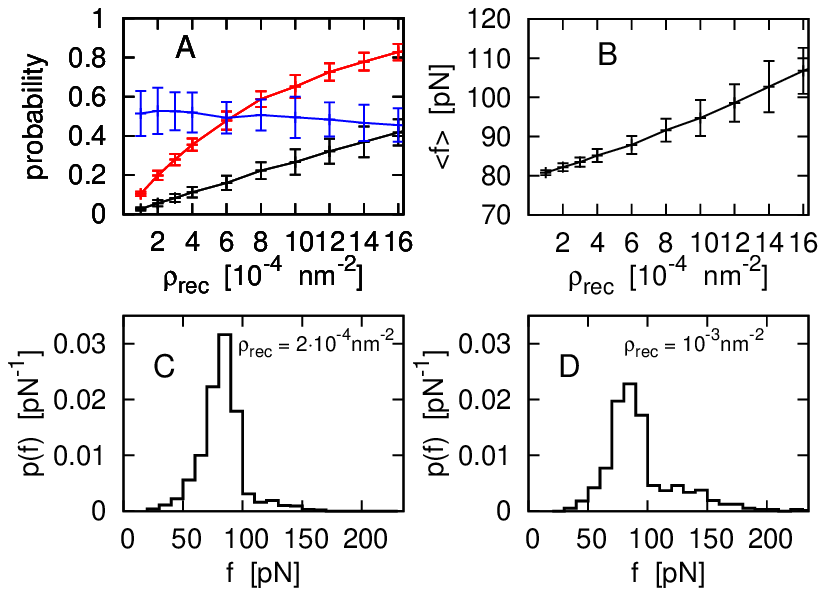} 
\caption{
}
\label{fig6}
\end{figure}
%%%%%%%%%%%%%%%%%%%%%%%%%%%%%%%%%%%%%%%%%%%%%%%%%%%%%%%%%%%%%%
\begin{figure}[t]
\epsfbox{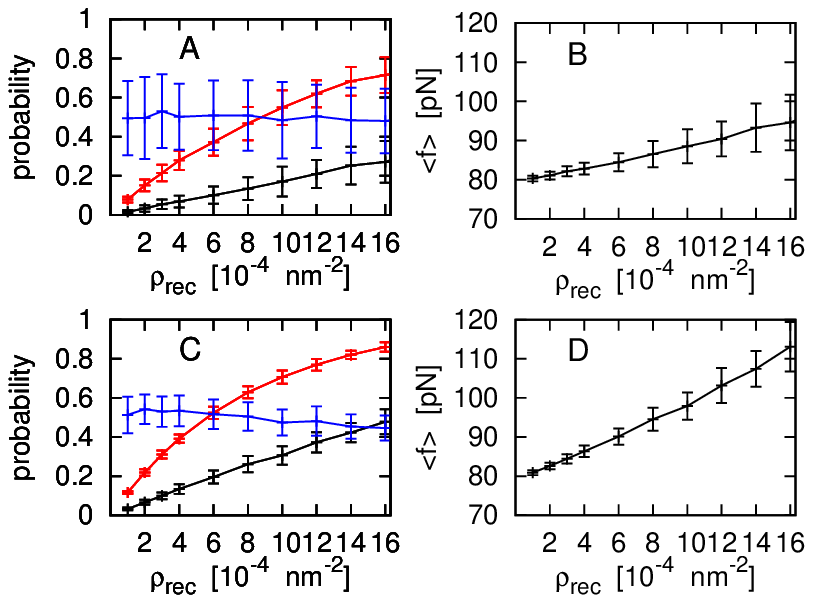} 
\caption{
}
\label{fig7}
\end{figure}
%%%%%%%%%%%%%%%%%%%%%%%%%%%%%%%%%%%%%%%%%%%%%%%%%%%%%%%%%%%%%%
\begin{figure}[t]
\epsfbox{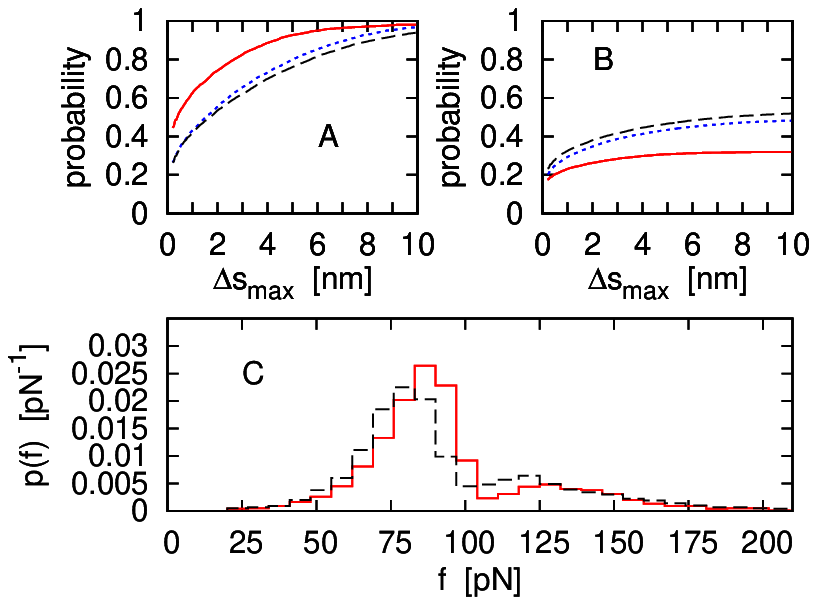} 
\caption{
}
\label{fig8}
\end{figure}
%%%%%%%%%%%%%%%%%%%%%%%%%%%%%%%%%%%%%%%%%%%%%%%%%%%%%%%%%%%%%%
\begin{figure}[t]
\epsfbox{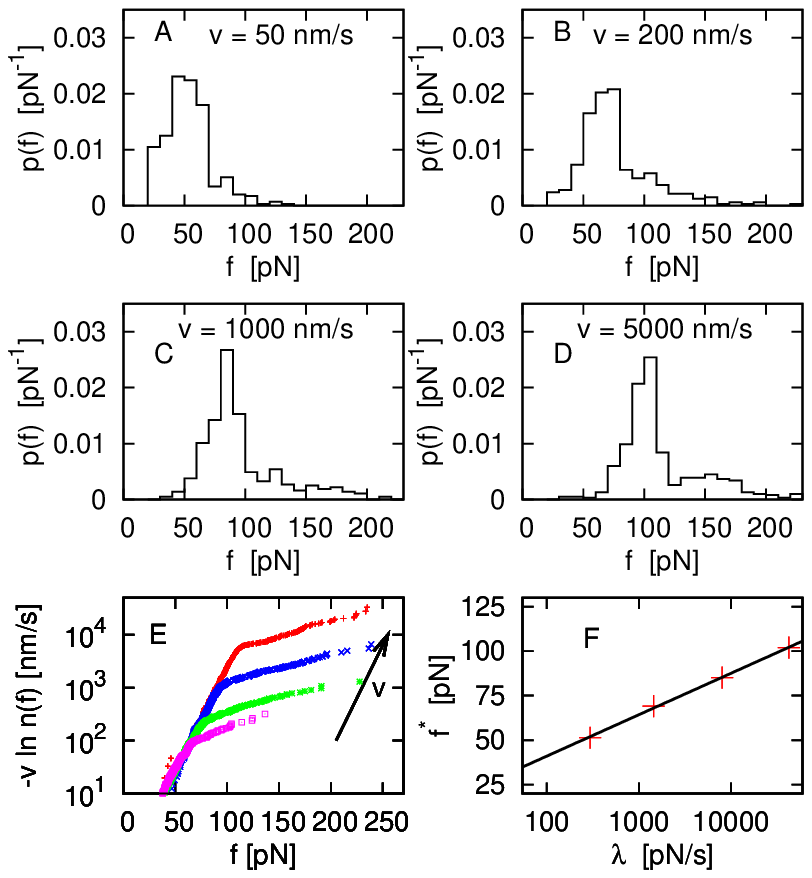} 
\caption{
}
\label{fig9}
\end{figure}

\end{document}